\def \beq {\begin{equation}}
\def \eeq {\end{equation}}
\def \beqa {\begin{eqnarray}}
\def \eeqa {\end{eqnarray}}

\newcommand{\req}[1]{(\ref{#1})}
\newcommand{\vect}[1]              % Format vector symbols as
           {\mbox{\boldmath$#1$}}  % italic boldface letters.
\documentclass[12pt]{article}
\usepackage{graphicx}

\makeatletter
\newcount\@tempcntc
\def\@citex[#1]#2{\if@filesw\immediate\write%
                  \@auxout{\string\citation{#2}}\fi
  \@tempcnta\z@\@tempcntb\m@ne\def\@citea{}\@cite{\@for\@citeb:=#2\do
    {\@ifundefined
       {b@\@citeb}{\@citeo\@tempcntb\m@ne\@citea%
                   \def\@citea{,}{\bf ?}\@warning
       {Citation `\@citeb' on page \thepage \space undefined}}%
    {\setbox\z@\hbox{\global\@tempcntc0\csname b@\@citeb%
                     \endcsname\relax}%
     \ifnum\@tempcntc=\z@ \@citeo\@tempcntb\m@ne
       \@citea\def\@citea{,}\hbox{\csname b@\@citeb\endcsname}%
     \else
      \advance\@tempcntb\@ne
      \ifnum\@tempcntb=\@tempcntc
      \else\advance\@tempcntb\m@ne\@citeo
      \@tempcnta\@tempcntc\@tempcntb\@tempcntc\fi\fi}}\@citeo}{#1}}
\def\@citeo{\ifnum\@tempcnta>\@tempcntb\else\@citea\def\@citea{,}%
  \ifnum\@tempcnta=\@tempcntb\the\@tempcnta\else
   {\advance\@tempcnta\@ne\ifnum\@tempcnta=\@tempcntb%
     \else \def\@citea{--}\fi
    \advance\@tempcnta\m@ne\the\@tempcnta\@citea\the\@tempcntb}\fi\fi}
\makeatother
\begin{document}
% >>>>>>>>>>>>>>>>>>>>>>>>>>>>>>>>>>>>>>>>>>>>>>>>>>>>>>>>>>>>>>>>>>>>
% TITLE AND AUTHORS.
%

\vspace*{-1.0cm}
\begin{center}
{\Large\bf Nuclei beyond the drip line}
\\[2.0cm]
{J.N. De $^*$, X. Vi\~nas,  S.K. Patra and M. Centelles}
\\[2mm]
{\it Departament d'Estructura i Constituents de la Mat\`eria,
     Facultat de F\'{\i}sica,
\\
     Universitat de Barcelona,
     Diagonal {\sl 647}, E-{\sl 08028} Barcelona, Spain}
\end{center}
% >>>>>>>>>>>>>>>>>>>>>>>>>>>>>>>>>>>>>>>>>>>>>>>>>>>>>>>>>>>>>>>>>>>>
% ABSTRACT, PACS.
%
\vspace*{2.0cm}
\begin{abstract}

In a Thomas-Fermi model, calculations are presented for 
nuclei beyond the nuclear drip line at zero temperature. These nuclei
are in equilibrium by the presence of an external gas, as may be envisaged
in the astrophysical scenario. We find that there is a limiting asymmetry 
beyond which these nuclei can no longer be made stable.

\end{abstract}

\mbox{}


\vfil 

%\hline
\noindent $^*$ On leave of absence from the Variable Energy
          Cyclotron Centre, 1/AF, Bidhannagar, Calcutta-700064,
          INDIA.

\pagebreak

% >>>>>>>>>>>>>>>>>>>>>>>>>>>>>>>>>>>>>>>>>>>>>>>>>>>>>>>>>>>>>>>>>>>>
% INTRO.
%
%\section{Introduction}

%%%%%%%%%%%%%%%%%%%%%%%%%%%%%%%%%%%%%%%%%%%%%%%%%%%%%%
%\setcounter{section}{1}
%\setcounter{equation}{0}
%\section*{1}
%%%%%%%%%%%%%%%%%%%%%%%%%%%%%%%%%%%%%%%%%%%%%%%%%%%%%%

Nuclear stability is limited by the drip lines. As one moves away 
from the valley of $\beta-$stability, e.g., by increasing the neutron excess, 
the neutron chemical potential ($\mu_n$) that is negative for stable nuclei 
steadily decreases in magnitude until it becomes zero at the
neutron drip line. Beyond this line the neutron chemical potential is
positive and the system cannot hold the excess neutrons together,
rendering itself unstable. Similar is the situation for nuclei
with a proton excess. It was, however, found that it is possible
to extend the drip lines ($\mu_n=0$ or $\mu_p=0$) provided the nuclei 
have a temperature.
In Ref. \cite{ref1} this was achieved by assuming the nucleus to be 
a liquid drop in
thermodynamical equilibrium (thermal and chemical) with a surrounding 
vapour which exerts an external pressure to stabilise the system,
whereas in Ref. \cite{ref2} it was done by assuming the nuclear drop to
be in metastable equilibrium (pressure equal to zero).

 In the  context of the equation of state (EOS) of cold dense matter, 
 equilibrium nuclei 
 far beyond the nominal drip lines are to be considered. There the nuclei are immersed 
in a nucleonic gas, mostly of neutrons, which exerts pressure to keep the
system stable. Nuclei in this scenario were first studied
by Langer et al \cite{lan} and later in a broader scope by Baym, Bethe and
Pethick \cite{ref10}. From a microscopic viewpoint there are two basic concerns.
 One, a thermodynamically consistent treatment of the coexistence \cite{ref4,ref5,ref6,pet}
 of the two phases of nuclear matter (namely, the nuclear liquid and the gas enclosing the liquid),
 the other, a plausible description of the interface between the liquid and the
 gas \cite{ref9,ref3}. As the density of the nuclear gas increases, the surface energy
 must decrease \cite{ref6}. In a recent paper, it was shown by Centelles et al \cite{ref3} that
 for semi-infinite nuclear matter this problem can be circumvented by solving for the
 density profile in the subtraction procedure \cite{ref11,ref12}, treating the density
 as the difference between its value in the nuclear-plus-gas phase and its value in the
 gas phase. The influence of the surrounding gas
 on the surface energy is then automatically taken into account. 
 They invoked a mean-field model and showed how asymmetric infinite
 or semi-infinite nuclear matter with large neutron excess (positive pressure, positive
 neutron chemical potential) can remain in stable equilibrium at zero temperature with
 a surrounding gas of drip  nucleons.

In this communication we generalise the idea of Ref. [10] to the case of finite
nuclei at  zero temperature and focus on the effect of the external gas on the structure and
stability of the nuclei with large neutron excess. We may stress that our main aim is
not to study the EOS of dense matter, but to give a better guide to understand 
the properties of isolated nuclei
immersed in a nucleonic gas. This knowledge may afterwards be taken as input for
a broader and careful study of the conditions under which nuclei can exist in neutron star matter. 
In order to treat the matter inside the nucleus with that outside in
a consistent fashion, we look for density solutions for the nucleus-plus-gas phase
and also for the gas phase alone and isolate the nucleus
from the surrounding gas by subtracting the latter from the former. This 
method  formally is  similar to that of finding solutions to the
coexistence between a hot nucleus and the evaporated nucleons around it
\cite{ref11,ref12}. We employ the Thomas-Fermi formalism, which has earlier been used as a 
convenient tool \cite{buc,oga,pi} to understand the EOS of dense nuclear matter. Here we resort
to the Thomas-Fermi method restructured for the subtraction procedure \cite{ref13} to get
the density solutions as mentioned earlier. 
For the nuclear force, we choose the
Skyrme interaction SkM$^*$ \cite{ref14}.
In doing the
calculations beyond the drip lines we find that the nuclear asymmetry $I$ cannot
be increased arbitrarily (we define the asymmetry of the finite nucleus as
$I=(N-Z)/A$). There is a limiting asymmetry beyond which the thermodynamic
equilibrium conditions cannot be met. 

The subtraction procedure in the Thomas-Fermi prescription
has been described in detail in Refs. \cite{ref3,ref13}; we present here
only the working equations. The method is based on the existence of two
solutions to the Thomas-Fermi equations, one corresponding to the
nucleus phase with the surrounding gas ($NG$) and the other corresponding to
the gas ($G$) alone. The densities in the $NG$ and the $G$ phase are
obtained by solving the coupled equations

\beq
(3\pi^2)^{2/3}\frac{\hbar^2}{2m_{\tau}^*}\rho_{NG}^{2/3} + V_{NG}^{\tau}
+V_{NG}^c(\rho_{NG},\rho_G)  =  \mu^{\tau}
\label{eqFN1}
%\nonumber \\[3mm]
%\\[3mm]
%& & \mbox{}
\eeq
\beq
(3\pi^2)^{2/3}\frac{\hbar^2}{2m_{\tau}^*}\rho_G^{2/3} + V_G^{\tau}
+V_G^c(\rho_{NG},\rho_G)=\mu^{\tau}.
\label{eqFN2}
\eeq
The density of the nucleus is then given by
\beq
\rho(r)=\rho_{NG}(r)-\rho_G(r)
\label{eqFN3}
\eeq
so that
\beq
A^{\tau}=\int \left [ \rho_{NG}^{\tau}(r) - \rho_G^{\tau}(r)\right ]
d^3r,
\label{eqFN4}
\eeq
where $A^{\tau}$ is the neutron or proton number ($N$ or $Z$) of the
nucleus in question. In Eqs. \req{eqFN1} and \req{eqFN2}, $\mu^{\tau}$ is
the chemical potential; it defines the nucleus as

\beqa
\mu^{\tau} & = & \frac{1}{A^{\tau}}\left \{\int \left [
(3\pi^2)^{2/3}\frac{\hbar^2}{2m_{\tau}^*}\rho_{NG}^{2/3}(r) + V_{NG}^{\tau}(r)
+V_{NG}^c(r)\right ]\rho_{NG}(r)d^3r \right .
\nonumber \\[3mm]
 & & \mbox{}
-
\left .
\int\left [(3\pi^2)^{2/3}\frac{\hbar^2}{2m_{\tau}^*}
\rho_G^{2/3}(r) + V_G^{\tau}(r)
+V_G^c(r)\right]\rho_G(r)d^3r\right \}.
\label{eqFN5}
\eeqa
Here $V^{\tau}$ are the nuclear part of the single-particle potential
for either the $NG$ or $G$ solution and $V^c$ is the Coulomb potential
coupling both the solutions;  $m_{\tau}^*$ used in Eqs. \req{eqFN1}
and \req{eqFN2} are the density-dependent
effective masses  for each phase.
The direct part of $V^c$ is the same for both phases and is given by
\beq
V_d^c(r)=\frac{e^2}{2}\int \left[\rho_{NG}^p(r')-
\rho_G^p(r')\right ] g(r,r')d^3r'
\label{eqFN6}
\eeq
with
\beq
g(r,r')=\left[(r+r')-\left | r-r'\right |\right]/(rr').
\label{eqFN7}
\eeq
The Coulomb potentials in the $NG$ and $G$ phase
differ only through the exchange terms:
$-e^2(3/\pi)^{1/3}(\rho^p_{NG,G})^{1/3}$.
They however have the same asymptotic value.
Equations \req{eqFN1} through \req{eqFN6} are solved self-consistently in
an iterative manner to yield the densities $\rho_{NG}(r)$ and
$\rho_G(r)$. In neutron-rich nuclei the proton density in the
gas phase is $\rho_G^p(r)=0$, similarly in 
proton-rich nuclei, there are no neutrons in the gas phase. Solutions obtained
in this manner ensure mechanical and chemical equilibrium between the nuclear
phase and the drip phase.

 We have done calculations for the isotopes of four systems, namely those 
of Ca, Zr, Sn and Pb. This covers a broad  range of charges.
As one adds more neutrons to a system with fixed $Z$, the
neutron chemical potential (which is negative for a stable nucleus) increases until it crosses
       zero meaning thereby that further addition of neutrons would not allow them to
       be part of a bound system causing them to drip (the neutron drip line). Similarly,
       removal of neutrons for a system with fixed charge increases the proton chemical potential
 for the system until it reaches the proton drip line when the system becomes
 unstable.

If these nuclei are, however, surrounded by a gas of nucleons as may well be
       the case in the astrophysical scenario, say after a supernova explosion, the excess
       pressure given by the gas (drip phase) may stabilise them and then one 
may reach the
       stability limits  beyond the nominal drip line. For example, $^{68}$Ca is the
       neutron drip-line nucleus in our calculation, but with a drip phase surrounding the
       nucleus one can reach $^{88}$Ca as a stable system. The density of the drip phase
       (in this case that of the drip neutrons) increases with the increasing number of
        neutrons in the nucleus. Similarly, $^{35}$Ca is the proton drip-line 
nucleus. One
       can, however, go further below in mass  to $^{25}$Ca with a surrounding proton gas. The drip phase of
       the protons is assumed to be charge neutral \cite{ref12}; in the
astrophysical context the
       proton gas is neutralised by the electrons pervading the gas. The external gas
       does not contribute to the Coulomb energy, the Coulomb potential $V^c$ is calculated
       from the subtracted proton density which is exactly the proton density of the
       isolated  nucleus.

           In Fig. 1, the density distributions of $^{340}$Pb are shown. The nucleus has sixty
       three more neutrons than the drip-line nucleus $^{277}$Pb as found in our calculation.
       We display the density of the nucleus-plus-gas and also the subtracted density. The
       subtracted density is found to be independent of the size of the box in which the calculations
       are done. The neutron drip-phase density is seen to grow to more
than one-tenth of
       the central neutron density. There are no protons in this drip phase. For comparison, the
       densities of the nucleus $^{208}$Pb are also displayed. One can easily discern the thicker
       neutron skin in $^{340}$Pb and also the effect of the asymmetry potential in pulling
       the proton distribution outwards from the origin as compared to that in $^{208}$Pb. 
The central proton density
is decreased to almost half of its value in $^{208}$Pb. In Fig. 2, the density distributions for the very 
       proton-rich nucleus $^{140}$Pb are shown. Here the drip phase consists only of protons. It is
       polarised because of the influence of the Coulomb field of the nucleus on the
       surrounding gas as seen in the case of evaporated protons from a hot nucleus 
       \cite{ref11,ref12}. In all cases, the nucleus-plus-gas solution
       coincides with the gas solution asymptotically at large distance for both neutrons and
       protons.

           In the calculations on infinite or semi-infinite nuclear matter in equilibrium
       with the drip phase, it was observed in Refs. \cite{ref4,ref9,ref3} that the 
asymmetry could be 
increased arbitrarily till the densities and asymmetries of both
       phases merged with each other. In the self-consistent calculations in finite systems,
       we however observe that one cannot add or remove neutrons from a nucleus arbitrarily.
       Beyond a certain point, further addition or removal creates instabilities in the
       system and no solution to the Thomas-Fermi problem can be found. We call this the limiting
       asymmetry. Establishing full chemical equilibrium between the nuclear phase and the
       drip phase becomes harder in these conditions, and beyond the limiting asymmetry it
       is no longer possible. This type of instabilities has a remarkable similarity to
       those found in the calculation of limiting temperatures for hot nuclei 
\cite{ref11,ref12}. As
       a function of the charge $Z$ of the nucleus,  the limiting asymmetry along with the asymmetry at the drip
       lines are plotted in Fig. 3. On the neutron-rich side, they are 
 found to be almost constant in the charge range we consider.
       On the neutron-deficient side, their magnitudes decrease with 
increasing charge. The wider gap between the nominal asymmetry (proton drip line) and the 
limiting asymmetry, due to the Coulomb repulsion is easily discernible.
In Fig. 4, the rms radii of neutrons and protons for the Ca and Pb isotopes are
displayed. One sees the growth of the neutron or proton skins with positive
or negative asymmetry. One further finds that except for the nuclei near the edges of
limiting asymmetry, the rms radii scale perfectly with $A$, the mass number.
On both sides of the mass scale, the sudden rise
in the neutron or proton radius points to the onset of instability.

   To conclude, we have done calculations in a suitably modified Thomas-Fermi model 
 for nuclei beyond the drip lines, when the nuclei are immersed in a nucleonic fluid.
 In such a prescription, no extra care is needed to treat the interface region
 of the nucleus and the surrounding fluid or to isolate the nucleus from the environment.
 From the numerical calculations, a limiting asymmetry beyond which nuclei even within the
 gaseous environment cannot exist is found; the delicate balance between the Coulomb force 
 and the diluted surface tension with increasing asymmetry and increasing density of the 
 environment very likely plays the crucial role there.
 We have not gone into the extensions of the Thomas-Fermi scheme, have
not taken shell effects into consideration, and have worked with the SKM* interaction
       whose validity at large asymmetry is not unquestionable. Sophistications in the
       approach or the use of different interactions may change the results somewhat 
quantitatively, but the qualitative
       features of the physics that emerge from our calculations, we believe,
       will remain unchanged.

The authors would like to acknowledge support from the DGICYT (Spain)
under grant PB98-1247 and from DGR (Catalonia) under grant
1998SGR-00011. J.N.D. and S.K.P. thank the Spanish  Ministry of Education for financial support 
with the grants SAB1999-0229 and SB97-OL174874  and the Departament d'Estructura i
Constituents de la Mat\`eria of the University of Barcelona for the kind
hospitality extended to them.

% REFERENCES.

%
\pagebreak

%
% >>>>>>>>>>>>>>>>>>>>>>>>>>>>>>>>>>>>>>>>>>>>>>>>>>>>>>>>>>>>>>>>>>>>
%
% FIGURE CAPTIONS.
%
\section*{Figure captions}
\begin{description}
\item[Figure 1.]
The total and subtracted neutron and proton densities for $^{340}$Pb are
shown in the upper panel. In this nucleus, there are no protons in the
drip phase. In the lower panel, the densities for $^{208}$Pb are displayed.
\item[Figure 2.]
The total and subtracted neutron and proton densities for the nucleus
$^{140}$Pb. In this nucleus, there are no neutrons in the drip phase.
In the nuclear interior (up to $\sim 10$ fm), the total and subtracted
proton densities are indistinguishable, beyond this, the polarisation of
the proton density in the drip phase is evident.
\item[Figure 3.]
The limiting and drip line asymmetry as a function of the charge number
of nuclei.
\item[Figure 4.]
The rms radii of neutrons and protons for the isotopes of Ca (upper panel) and
Pb (lower panel). The open circles and squares refer to the calculated neutron
and proton radii as indicated. The solid and the  dashed lines refer to 
linear fits of the radii to the mass number $A$ (the nuclei at the edges have
been excluded from the fit).
\end{description}
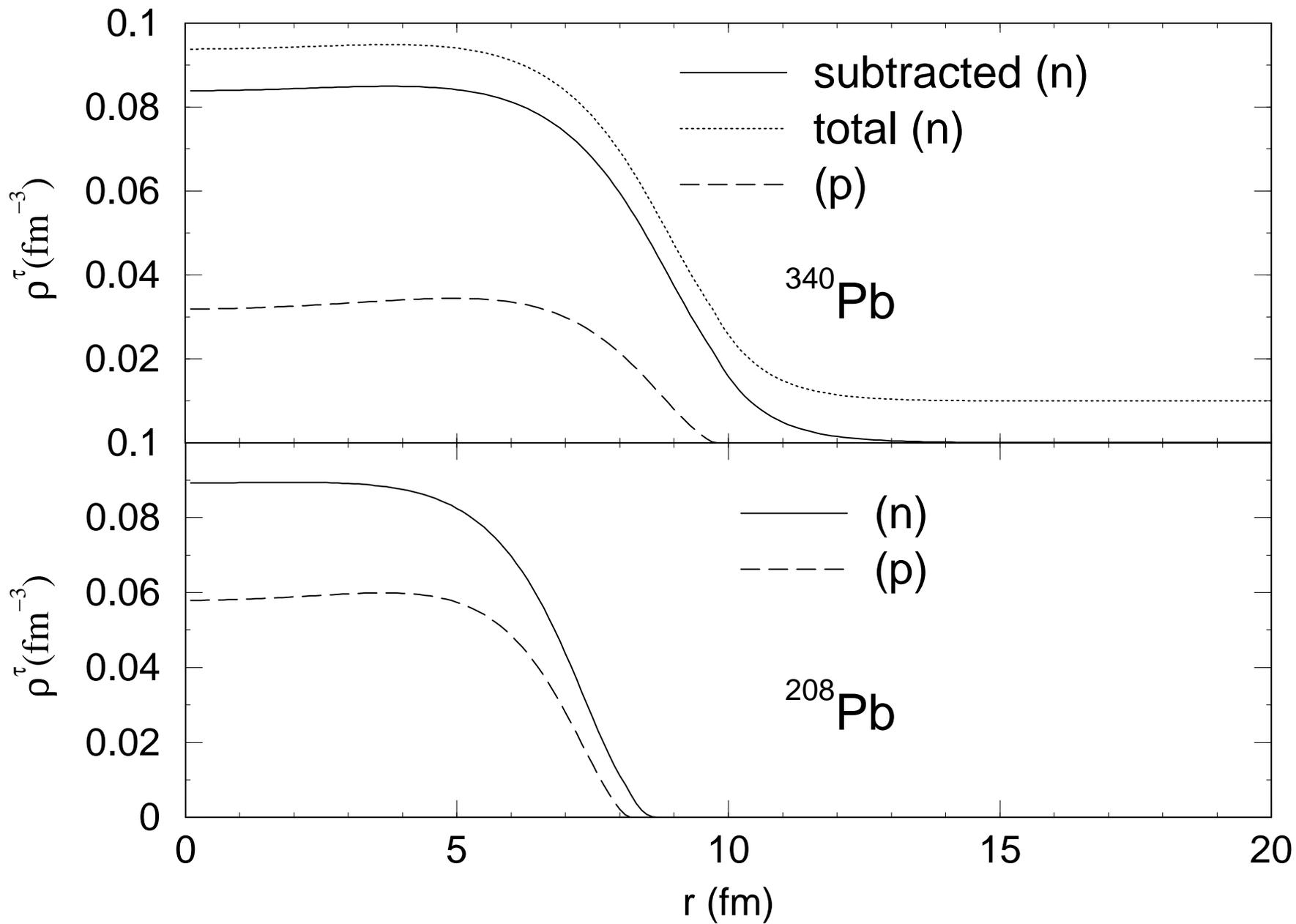
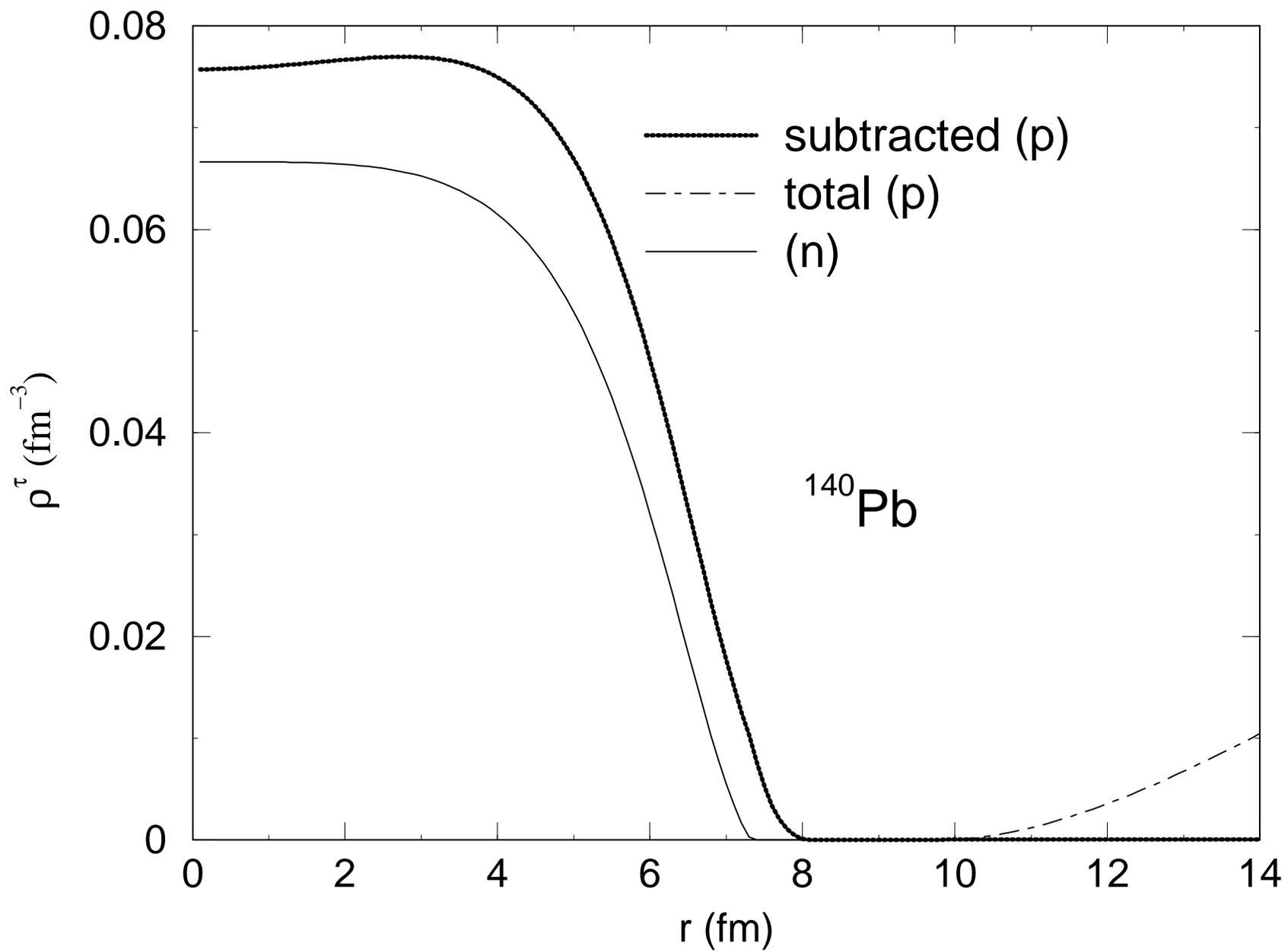
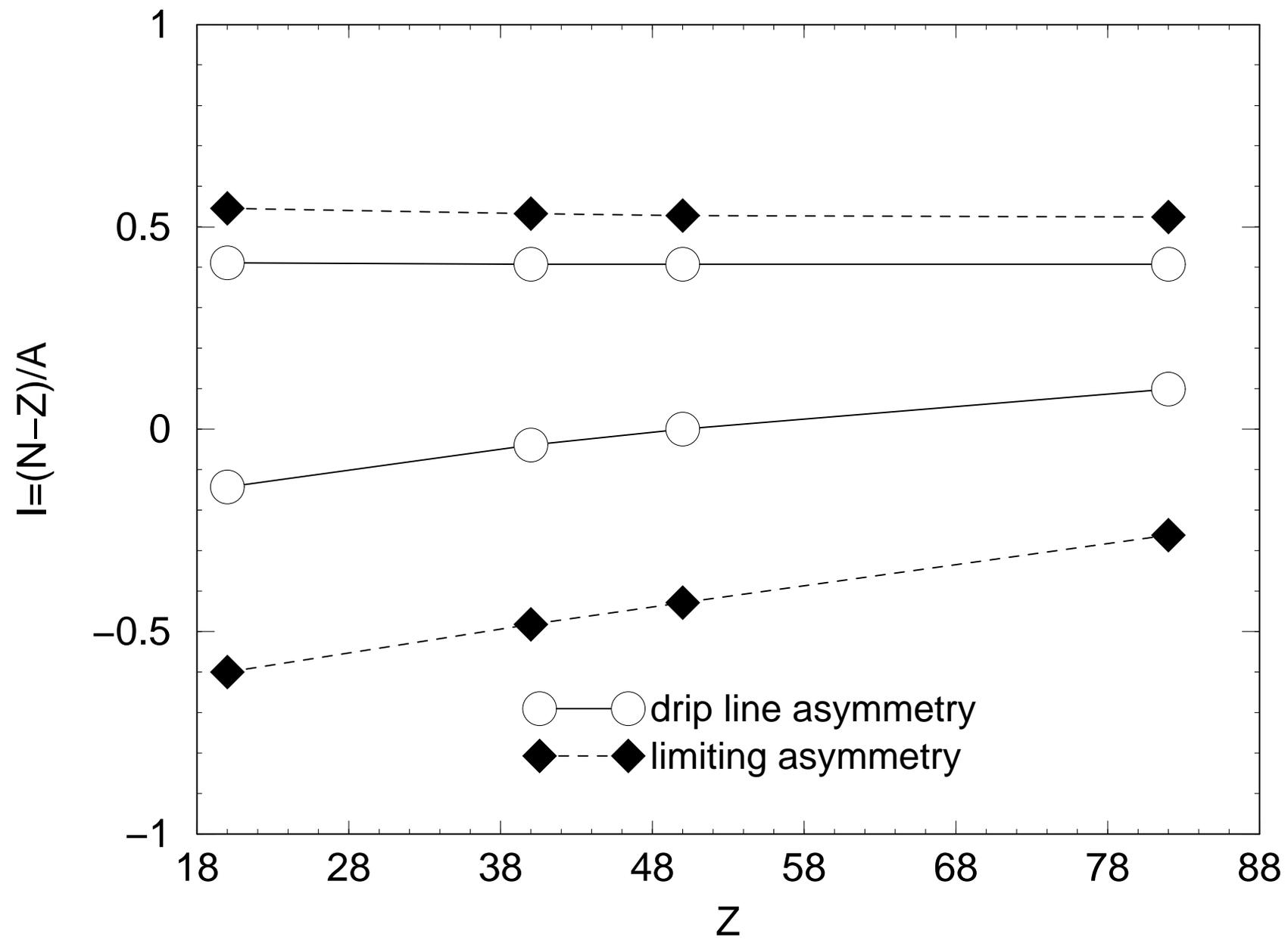
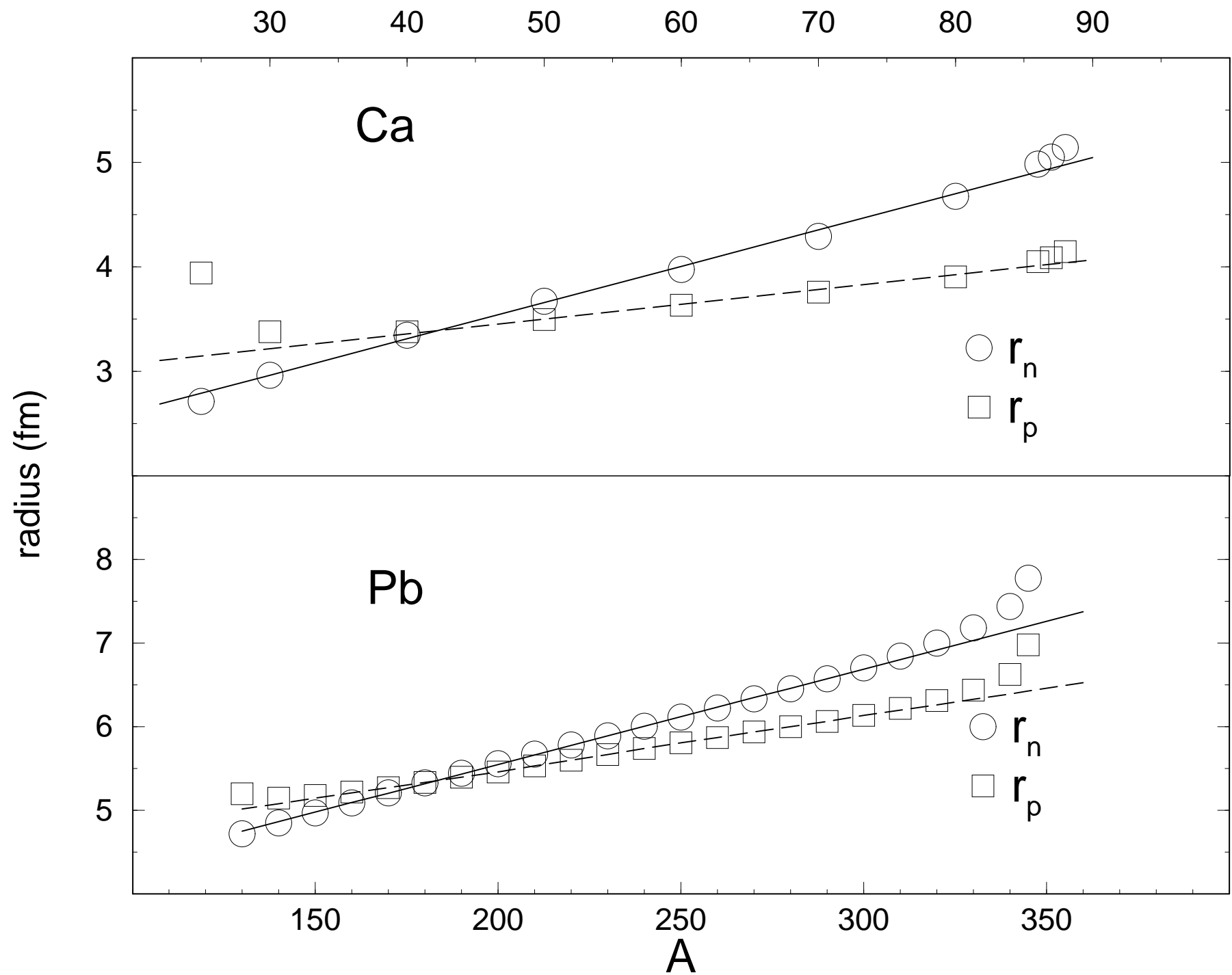
\end{document}